\documentclass[10pt,prl,aps,twocolumn,showpacs,floatfix]{revtex4}
\usepackage{graphicx}
\usepackage{amsmath}
\usepackage{amssymb}
\usepackage{bm} % bold math
\usepackage{color}
\usepackage{braket} 

\begin{document}
\draft
\title{Multiple universalities in order-disorder magnetic phase transitions}
\author{H. D. Scammell and O. P. Sushkov }
\affiliation{School of Physics, The University of New South Wales,
  Sydney, NSW 2052, Australia}
\date{\today}
\begin{abstract}
Phase transitions in isotropic quantum antiferromagnets are associated with the condensation of bosonic triplet excitations. In three dimensional quantum antiferromagnets, such as TlCuCl$_3$, condensation can be either pressure or magnetic field induced. The corresponding magnetic order obeys universal scaling with thermal critical exponent $\phi$. Employing a relativistic quantum field theory, the present work predicts the emergence of multiple (three) universalities under combined pressure and field tuning. Changes of universality are signalled by changes of the critical exponent $\phi$. 
Explicitly, we predict the existence of two new exponents $\phi=1$ and $1/2$ as well as recovering the known exponent $\phi=3/2$. We also predict logarithmic corrections to the power law scaling.

\end{abstract}

\maketitle

\begin{figure}[t!]
 {\includegraphics[width=0.2385\textwidth,clip]{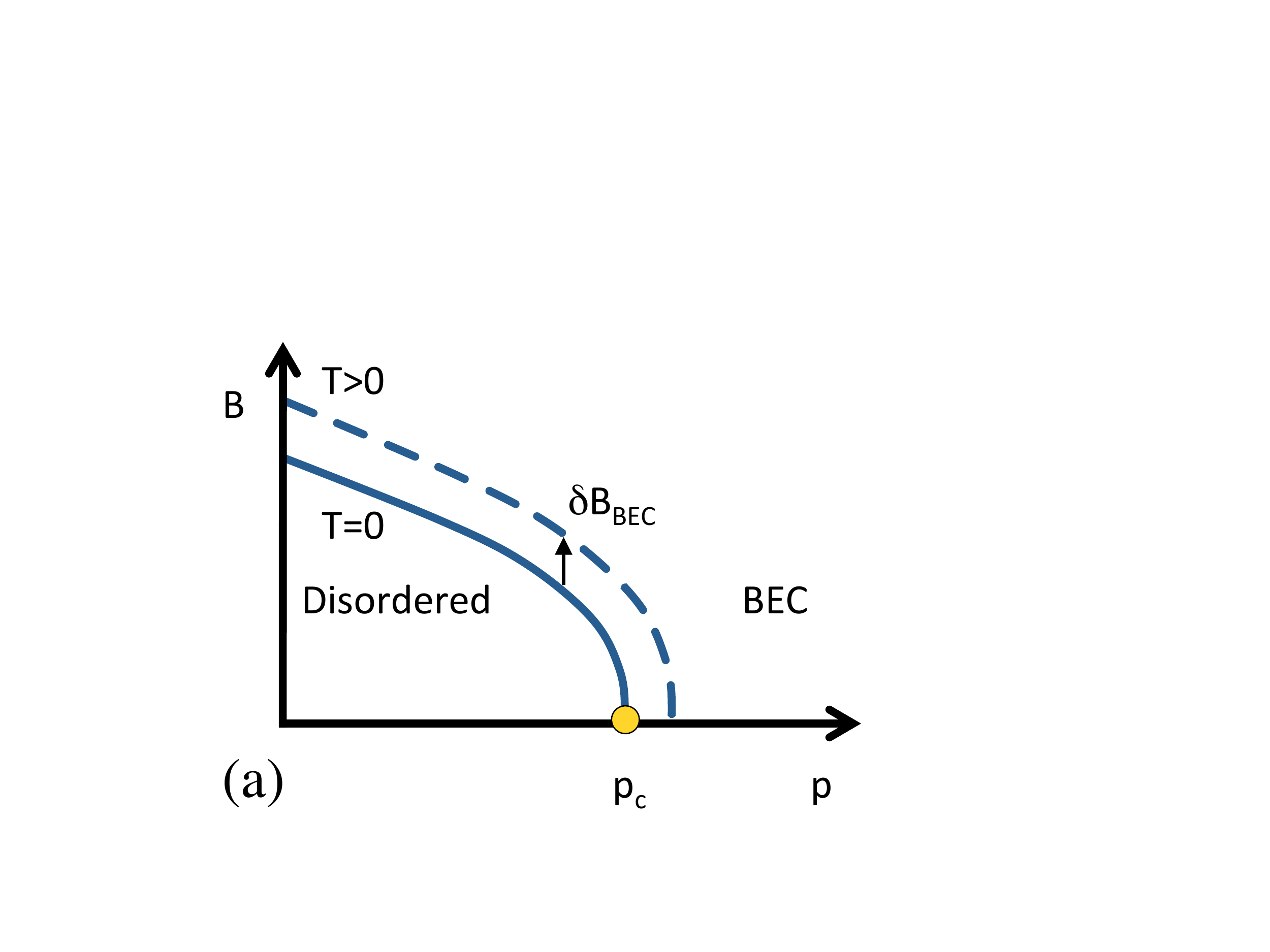}}
  {\includegraphics[width=0.2385\textwidth,clip]{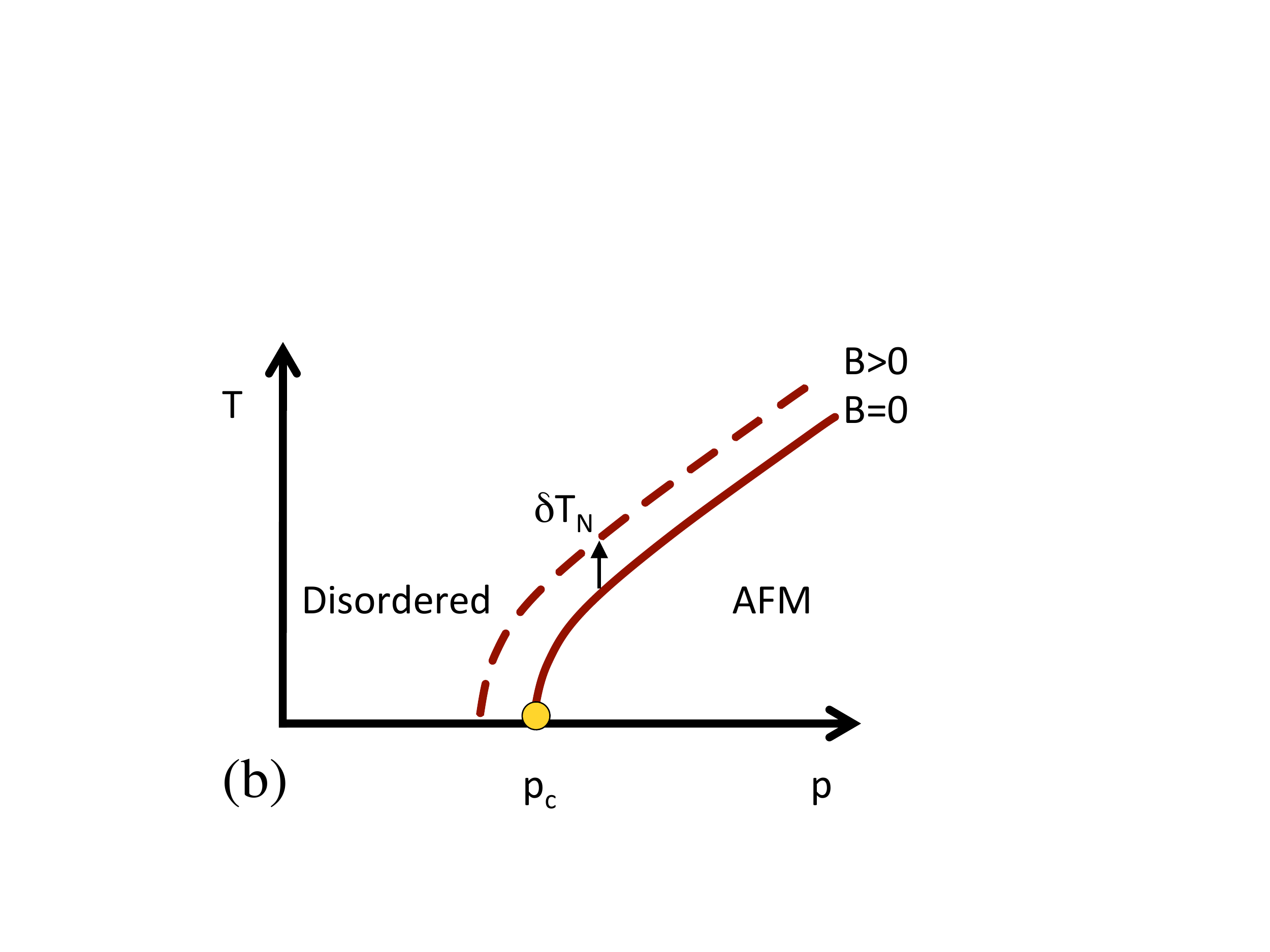}}
\caption{  Critical field and temperature power law shifts. (a) Shift of critical field-pressure line with temperature $\delta B_{BEC}\sim T^{\phi}$. Solid blue curve is at zero temperature, dashed blue at non-zero temperature. (b) Shift of critical (N\'eel) temperature-pressure line with field $\delta T_N\sim B^{1/\phi}$. Solid red curve is at zero field, dashed red at non-zero field.  }
\label{shifts}
\end{figure}

\begin{figure}[h!]
 {\includegraphics[width=0.35\textwidth,clip]{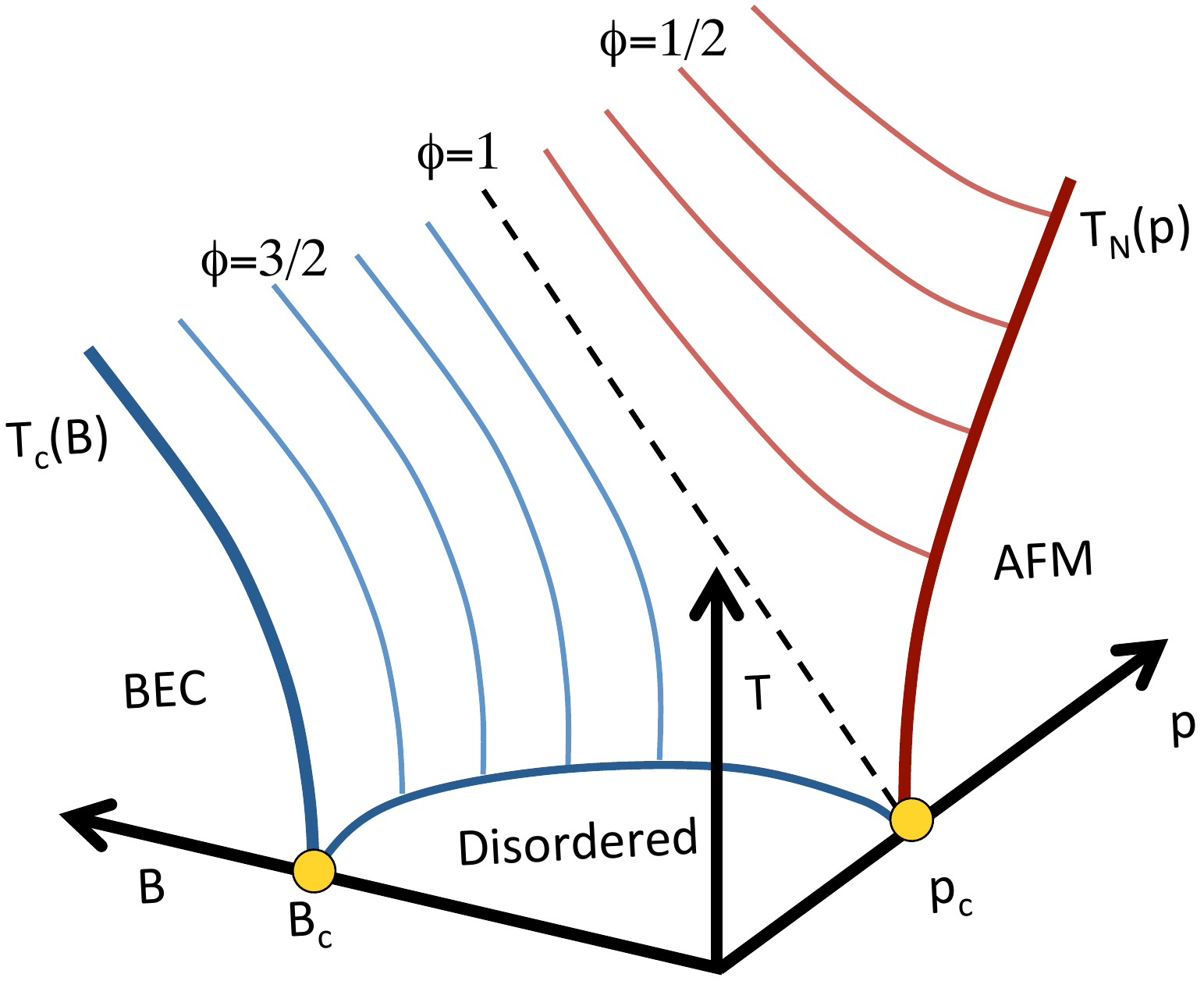}}
\caption{ Multiple universalities in the $(p, B, T)$ phase diagram. Blue curves correspond to the BEC transition lines; here $p<p_c$ and the critical exponent is $\phi=3/2$. Red curves correspond to the N\'eel transition lines; here $p>p_c$ and the critical exponent is $\phi=1/2$. The dashed, black curve shows the critical pressure transition line, with critical exponent $\phi=1$.
 }
\label{phase}
\end{figure}

Pressure and magnetic field induced condensate phases in quantum magnetic systems have become instrumental to our understanding of universal, critical-phenomena. 
A great effort (experimental, numerical and theoretical) has been devoted to uncovering and categorising the universal features of critical magnetic condensate phases. 
The present work considers three dimensional (3D) quantum antiferromagnets (QAF), where the combined interplay between pressure, magnetic field and temperature $(p, B, T)$ remains theoretically unexplored, yet offers an exiting arena for theorists and experimentalists alike to uncover new universal behaviour. In Figure \ref{shifts} we present the generic phase diagrams of dimerised QAFs such as TlCuCl$_3$, KCuCl$_3$, and CsFeCl$_3$. Panel (a) shows the magnon Bose condensation (BEC) line in the field-pressure diagram, and panel (b) shows the antiferromagnetic (AFM) transition line in the temperature-pressure diagram. It is also instructive to look at Figure \ref{phase} which shows the 3D $(p,B,T)$ phase diagram.
The point of primary interest is the critical field-critical temperature power law, 
\begin{align}
\label{PowerLaw}
{\text a}: \delta B_{BEC}&\sim T^{\phi},  && {\text b}: \delta T_N\sim B^{1/\phi},
\end{align}
The shift of the BEC transition line at small temperature is shown schematically in Fig. \ref{shifts}a; while the shift of the AFM/N\'eel transition line at small field is in Fig. \ref{shifts}b. 

It is widely believed that at $p<p_c$, $\phi=3/2$ is the universal BEC exponent, which can be obtained from the scaling arguments on the dilute Bose gas \cite{Fisher, Uzunov} or explicitly for magnon BEC \cite{Giarmarchi, Nikuni}. For a review see \cite{BECreview}. On the other hand, experiment (on TlCuCl$_3$ and KCuCl$_3$ \cite{Shiramura,Kato,Ishii,TanakaBc}) and numerics \cite{Wessel} show $1.5\leq\phi\lesssim2.3$, depending crucially on which temperature range is used for fitting \cite{BECreview, NohadaniQMC}. We understand recent data on 3D QAF CsFeCl$_3$ \cite{CsFeCl3}, taken along the thick blue-red solid lines in Fig. \ref{phase}, as a hint for a significant and unexpected evolution of the index $\phi$ along the line.
%%%%%%%%%%%%%%%%
%%%%%%%%%%%%%%%%%%%%%%%%%%%%
%%%%%%%%%%%%%%%%
%%%%%%%%%%%%%%%%%%%%%%%%%%%%

The primary goal of the present work is to derive the evolution of the critical index $\phi$ across the phase diagram. Another goal is to explain why the index depends on the fitting range; even if {\it a priori} the range seems to be very narrow. We will show that answers to both questions are related to the quantum critical point $(p,B,T)=(p_c,0,0)$. Ultimately, the quantum critical point (QCP) governs the evolution of the critical index $\phi$ across the phase diagram. This is illustrated in Fig. \ref{phase}.

Previous theoretical approaches were concentrated at the BEC transition, $p<p_c$. They employed a dilute Bose gas model \cite{Nikuni, BECreview}  and/or bond-operator technique \cite{Sirker}. In the end, these techniques rely on the Hartree-Fock-Popov approximation, yet it is known that the Hartree-Fock-Popov approximation breaks down in the vicinity of a critical point \cite{HFPAreview}. In the present work we employ a quantum field theory approach which naturally describes quantum critical points.

The quantum phase transition (QPT) between ordered and disordered phases
is described by the effective field theory with the following Lagrangian
\cite{Sachdev11,Kulik},
\begin{align}   
\label{Lagrangian}
{\cal L}&=\frac{1}{2}(\partial_{t}{\vec{\varphi}}-\vec{\varphi}\times\vec{B})^2-\frac{1}{2}(\vec{\nabla}{\vec{\varphi}})^2-\frac{1}{2}m^2_0{\vec{\varphi}}^{\ 2}-\frac{1}{4}\alpha_0\vec{\varphi}^{\ 4}.
\end{align}
The vector field $\vec{\varphi}$ describes staggered magnetisation, $B$ is an external applied field, and for now we set $g\mu_B = 1$. We now briefly outline the {\it mean-field} phase transitions captured by this Lagrangian. Consider first $B=0$, the pressure induced QPT results from tuning the mass term, $m_0^2$, for which we take the 
linear expansion $m^2_0(p)=\gamma^2(p_c-p)$, where $\gamma^2>0$ is a 
coefficient and $p$ is the applied pressure. Varying the pressure leads 
to two distinct phases; (i) for $p<p_c$ we have $m^2_0>0$, and the classical 
expectation value of the field is zero $\varphi_c^2=0$. This describes the 
magnetically disordered phase, 
the system has a global O(3) rotational symmetry, and the excitations are 
gapped and triply degenerate. (ii) For pressures $p>p_c$ we have $m^2_0<0$, 
and the field obtains a non-zero classical expectation value 
$\varphi^2_c=\frac{|m^2_0|}{\alpha_0}$.
This describes the magnetically ordered, antiferromagnetic phase. 
Varying $m^2_0$ from positive to negative spontaneously breaks the O(3) 
symmetry  of the system.

\begin{figure}[t!]
 {\includegraphics[width=0.238\textwidth,clip]{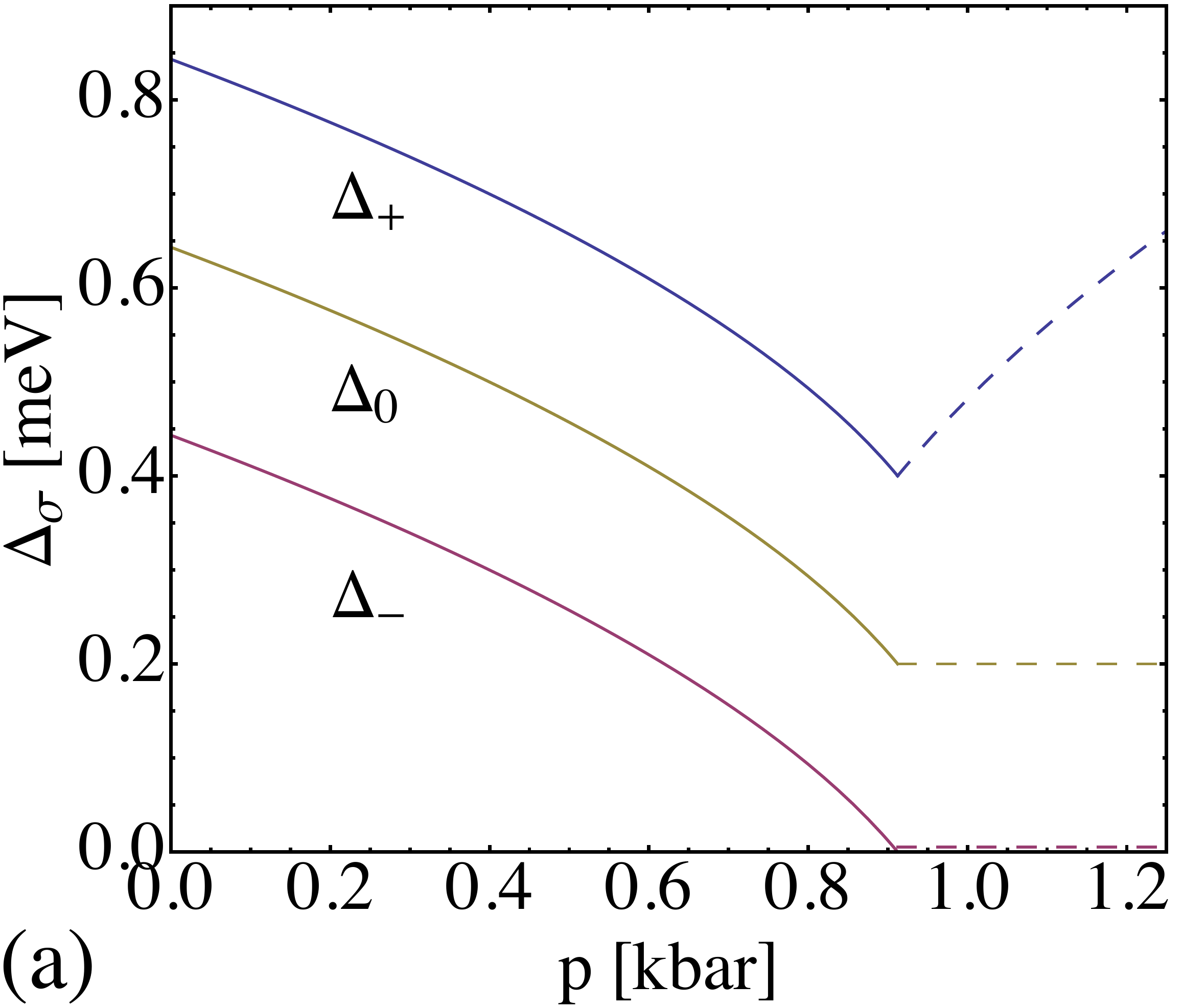}}
 {\includegraphics[width=0.238\textwidth,clip]{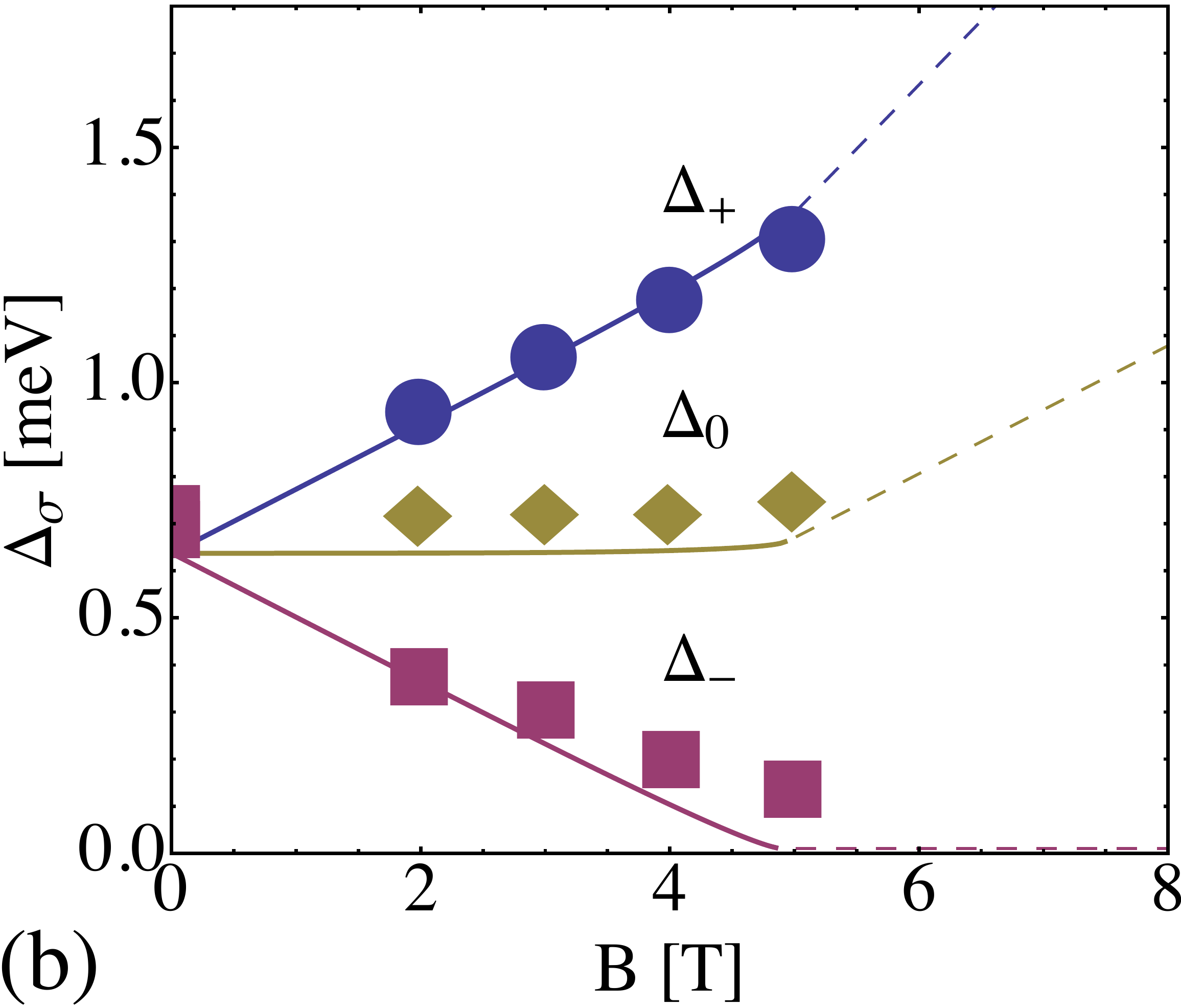}}
\caption{ Excitation gaps $\Delta_\sigma$: (Left) pressure driven at fixed field $B=0.2$ meV and $T=0$. (Right) field driven at $p=0$ kbar and $T=1.5$ K. Solid lines are theoretical results derived in this paper. Markers indicate experimental data for TlCuCl$_3$  \cite{RueggGaps,RueggGaps2}. 
 }
\label{Gaps}
\end{figure}

Next consider non-zero $B$ at fixed $p<p_c$: For $B<B_c=m_0$ the system has O(2) symmetry, and the degeneracy of the triplet modes is lifted by Zeeman splitting. The field induced QPT results from tuning $B>m_0$. The condensate field is $\varphi^2_c=\frac{B^2-m_0^2}{\alpha_0}$. To determine the order-disorder (BEC or AFM) transition line one can approach the transition starting from either the ordered or disordered phase. In this work we start from the latter; all results are derived starting from disordered phase. There are three magnetic excitations with ladder polarisation $\sigma=-,0,+$. The polarisation is the projection of angular momentum on the direction of magnetic field. In Figure \ref{Gaps} we summarise the results for the evolution of the three mode gaps through the field and pressure quantum phase transitions, separately. Explicit parameters correspond to those found in Ref. \cite{Asymptotic} for TlCuCl$_3$.  Here we disregard the small easy-plane anisotropy seen in TlCuCl$_3$, which has been shown to have negligible influence on the critical properties \cite{Asymptotic}, see also comment \cite{comment}.
%%%%%%%%%%%%%%%%
%%%%%%%%%%%%%%%%%%%%%%%%%%%%

{\it Beyond mean-field}:
Everywhere in the text $m_0^2=\gamma_0^2(p_c-p)$ and $\alpha_0$ represent the zero temperature mass tuning 
parameter and coupling constant without quantum fluctuation corrections. Taking into account quantum and thermal fluctuation corrections due to interaction term 
$\frac{1}{4}\alpha_0\vec{\varphi}^{\ 4}$, we will denote the renormalized parameters $m_0^2 \to m_{\Lambda,\sigma}^2$ and $\alpha_0\to\alpha_\Lambda$. The explicit form for $m_{\Lambda,\sigma}^2=m_{\Lambda,\sigma}^2(p,T,B)$ depends on the location within the phase diagram, and polarisation $\sigma$. Full details are presented in Supplementary Material  C and D, while expressions are presented below. The strength of the coupling $\alpha_\Lambda$ determines the strength of all interactions in the theory, and is dependent on the energy scale $\Lambda$. Generically, the one-loop renormalized coupling takes the form \cite{Zinn, Asymptotic}
\begin{align}
\label{coupling}
 \alpha_\Lambda=\frac{\alpha_0}{1+11\alpha_0/(8\pi^2)\ln(\Lambda_0/\Lambda)}.
 \end{align} 
 Specifically for the problem at hand, the coupling runs with scale $\Lambda=\max\{m_{\Lambda,\sigma}, B, T\}$. Accordingly, there is just a single point on the phase diagram at which all energy scales vanish $\Lambda\to0$; the quantum critical point $(p_c,0,0)$, see Fig. \ref{phase}. At this point the coupling runs to zero $\alpha_\Lambda\to0$ (asymptotic freedom). The running of the coupling constant will play an essential role in resolving our main goals/questions: Why the index $\phi$ depends on the location within the phase diagram, and; why the expected index $\phi=3/2$ in the BEC regime depends on the fitting range. 

 In the disordered phase the Euler-Lagrange equation with (\ref{Lagrangian}) results in the following dispersion
\begin{align}
\label{Zeeman}
\omega_{k}^\sigma&=\sqrt{k^2+m_{\Lambda,\sigma}^2}+\sigma B.
\end{align}
where $m_{\Lambda,\sigma}$ is the renormalised mass. Note that the $\sigma B$ term is not renormalised. This is a consequence of a Ward identity (Larmor theorem). While the stationary states (\ref{Zeeman}) have a fixed ladder polarisation, technically it is more convenient to calculate fluctuation corrections in the Cartesian basis $\vec{\varphi}=(\varphi_x,\varphi_y,\varphi_z)$. Let us denote by  ${\cal V}$ the part of the Lagrangian (\ref{Lagrangian}) independent of derivatives. Then, using a Wick decoupling of the interaction term $\frac{1}{4}\alpha_0\vec{\varphi}^{\ 4}$, in the single loop approximation we find
\begin{align}
\notag\frac{\partial^2{\cal V}}{\partial\varphi_x^2}&=m_0^2-B^2+3\alpha_0\braket{\varphi_x^2} + \alpha_0\braket{\varphi_y^2} +\alpha_0\braket{\varphi_z^2}\\
\notag\frac{\partial^2{\cal V}}{\partial\varphi_y^2}&=m_0^2-B^2+\alpha_0\braket{\varphi_x^2} + 3\alpha_0\braket{\varphi_y^2} +\alpha_0\braket{\varphi_z^2}\\
\label{curveRG}
\frac{\partial^2{\cal V}}{\partial\varphi_z^2}&=m_0^2+\alpha_0\braket{\varphi_x^2} + \alpha_0\braket{\varphi_y^2} +3\alpha_0\braket{\varphi_z^2}
\end{align}
where $\braket{\varphi_x^2}$ is the loop integral over the Green's function of field $\varphi_x$. An explicit calculation shows $\braket{\varphi_x^2}=\braket{\varphi_y^2}$, hence from equations (\ref{curveRG}), we have rather trivially satisfied the O(2) Ward identity: $\partial^2{\cal V}/\partial\varphi_x^2-\partial^2{\cal V}/\partial\varphi_y^2=0$. Further details are presented in Supplementary Material  A and B.

Quantum corrections corresponding to (\ref{curveRG}) come from the scale $\Lambda<q<\Lambda_0$. Hence they must be accounted via single loop renormalization group (RG). The thermal part of (\ref{curveRG}) comes from $q\sim T$, hence here the simple single loop approximation is sufficient. All in all, calculations presented in Supplementary Material D give
\begin{align}
\label{massRG}
\notag \frac{\partial^2{\cal V}}{\partial\varphi_i^2}&=m_{\Lambda,\pm}^2(T)-B^2\\
\frac{\partial^2{\cal V}}{\partial\varphi_z^2}&=m_{\Lambda,0}^2(T)
\end{align}
where $\varphi_i=\{\varphi_x,\varphi_y\}$, and  the renormalised masses are
\begin{align}
\notag m_{\Lambda,\pm}^2&\hspace{-0.05cm}=\hspace{-0.05cm}m_0^2\left[\frac{\alpha_\Lambda}{\alpha_0}\right]^{\frac{5}{11}}\hspace{-0.2cm}+\Sigma_T\\
\label{mR}
\notag m_{\Lambda,0}^2&\hspace{-0.05cm}=\hspace{-0.05cm}m_0^2\left[\frac{\alpha_\Lambda}{\alpha_0}\right]^{\frac{5}{11}}\hspace{-0.2cm}+\hspace{-0.05cm}\alpha_\Lambda\hspace{-0.05cm} \sum_k 1/\omega_k^0\hspace{-0.05cm}\{ \hspace{-0.05cm}n(\omega_{{\bm k}}^{+})\hspace{-0.05cm}+\hspace{-0.05cm}n(\omega_{{\bm k}}^{-})\hspace{-0.05cm}+\hspace{-0.05cm}3n(\omega_{{\bm k}}^{0})\hspace{-0.05cm}\}\\
\Sigma_T&\hspace{-0.05cm}\equiv\alpha_\Lambda\sum_k 1/\omega_k^0\{ 2n(\omega_{{\bm k}}^{+})+2n(\omega_{{\bm k}}^{-})+n(\omega_{{\bm k}}^{0})\}.
\end{align}
Here $n({\omega}_{\bm k})=1/(e^{\frac{\omega_{\bm k}}{T}}-1)$,
and we introduce the function $\Sigma_T$ for brevity. Obviously, expansions of Eq.'s (\ref{mR}) in powers of $B$ contain only even powers. Interestingly these expansions are different for $m_{\Lambda,\pm}$ and $m_{\Lambda,0}$. Therefore the relation $\omega_{\bm k}^+-\omega_{\bm k}^0=\omega_{\bm k}^0-\omega_{\bm k}^-$, which is exact at $T=0$, does not hold at non-zero $T$. At non-zero $T$ the relation is valid only up to the linear in $B$ approximation.  

In a magnetic field, the condition of condensation follows from Eq. (\ref{Zeeman}), $m_{\Lambda,\pm}-B_c=0$. Using (\ref{mR}) this equation can be rewritten as 
\begin{align}
\label{criticalTemp}
\Sigma_T=B_c^2-m_0^2\left[\frac{\alpha_\Lambda}{\alpha_0}\right]^{\frac{5}{11}}.
\end{align}
There are three distinct cases: (I) Above the critical pressure, when $T_c=T_N$ {\it i.e.} critical temperature equals the AFM/N\'eel temperature; (II) exactly at the critical pressure, $p=p_c$; (III) below the critical pressure, when $T_c=T_{BEC}$. At zero magnetic field, the critical temperature in case (I), Eq. (\ref{criticalTemp}), is identical to the equation for the N\'eel temperature derived in Ref. \cite{Asymptotic}.

Consider case (I); $p>p_c$. In this case according to Eq. (\ref{PowerLaw}b) the N\'eel temperature varies in a weak magnetic field. To calculate $\Sigma_T$ at $B\to0$ we take the critical line dispersions $\omega_{\bm k}^+=\omega_{\bm k}^-=\omega_{\bm k}^0=k$. Hence $\Sigma_T=\frac{5\alpha_{\Lambda}}{12} T^2$, where $T=T_{N0}+\delta T_N$; $T_{N0}$ is the N\'eel temperature in zero magnetic field. Hence using Eq.(\ref{criticalTemp}) we find
\begin{align}
\label{I}
{\text (I):}\hspace{0.2cm} \delta T_N&=\frac{6}{5\alpha_{\Lambda}}\frac{B^2}{T_{N0}} &&{\text at}\hspace{0.2cm} B\ll T_{N0}.
\end{align}
So the critical index in Eq. (\ref{PowerLaw}b) is $\phi=1/2$. 

\begin{figure}[t!]
{\includegraphics[width=0.31\textwidth,clip]{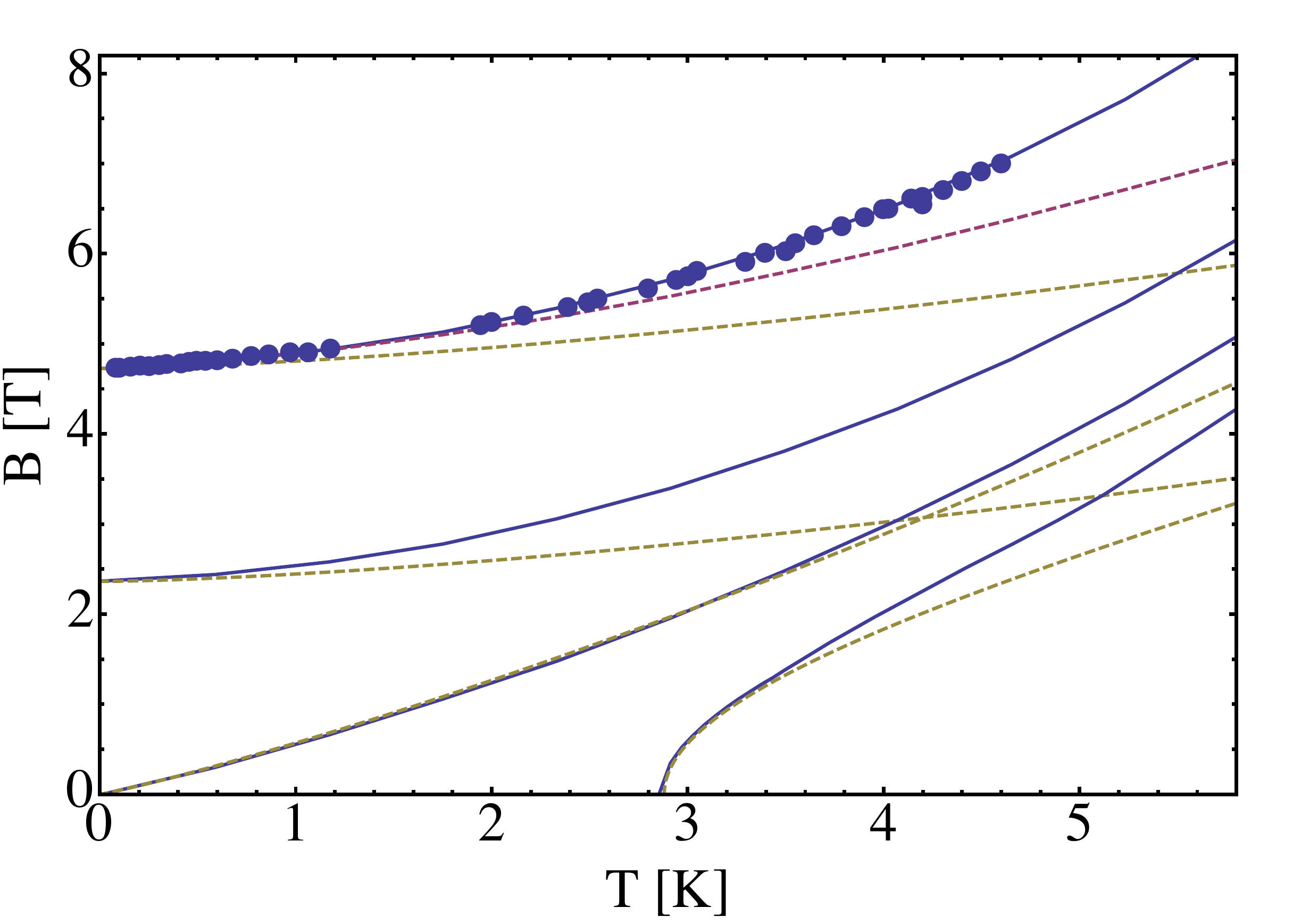}}
\caption{Critical field vs temperature: Dashed yellow curves show solutions to scaling equations (\ref{I}), (\ref{II}) and (\ref{III}). Dashed maroon shows solution of (\ref{criticalTemp}) that accounts for thermal mixing of non-critical modes, but does not account for running coupling; coupling is at fixed value $\alpha_\Lambda\to\alpha_{\Delta_0}=0.169\times 8\pi$.  Solid blue lines are the solution to (\ref{criticalTemp}) with full account of non-critical modes and logarithmic running coupling. Blue points are experimental data from \cite{TanakaBc, Oosawa, Shindo}.
 }
\label{BcT}
\end{figure}

In Ref. \cite{Asymptotic} the set of parameters describing TlCuCl$_3$ was determined
\begin{align}
\label{params}
\notag p_c&=1.01\ \text{kbar}, \  \gamma = 0.68\ \text{meV/kbar$^{1/2}$}, \\
\frac{\alpha_0}{8\pi} &=0.23, \ \ \ \ \ \ \Lambda_0=1\ \text{meV}.
\end{align}
When fitting experimental data in Ref. \cite{Asymptotic} the thermal line-broadening had been accounted via $\omega=k\to\omega=\sqrt{k^2+\xi^2 T^2}$, $\xi=0.15$. Therefore, if we use the set of parameters (\ref{params}) to determine the value of the running coupling constant $\alpha_\Lambda$, Eq. (\ref{coupling}), the coefficient in (\ref{I}) has to be corrected accordingly; $\frac{6}{5\alpha_{\Lambda}}\to1.14 \frac{6}{5\alpha_{\Lambda}}$. In Fig.\ref{BcT} we illustrate Eq.(9) by dashed yellow line originating from 
$T_{N0}=2.8$K. The couling constant is $\alpha_{\Lambda}/8\pi=\alpha_{T_{N0}}/8\pi=0.107$.
For comparison, the solid blue line originating from 2.8K represents exact solution of 
Eq.(8) with coupling constant running along the line.

Consider case (II); tuning exactly to the quantum critical point, $p=p_c$, $T_{N0}=0$. Again, to calculate $\Sigma_T$ at $B\to0$ we have to take the critical line dispersions $\omega_{\bm k}^+=\omega_{\bm k}^-=\omega_{\bm k}^0=k$ and hence again $\Sigma_T=\frac{5\alpha_{\Lambda}}{12} T^2$. Substitution into (\ref{criticalTemp}) gives
\begin{align}
\label{II}
{\text (II):}\hspace{0.2cm} B_c&=\sqrt{\frac{5\alpha_{\Lambda}}{12}}T&&{\text at}\hspace{0.2cm} B_c\ll T.
\end{align}
The condition $B_c\ll T$ is satisfied at sufficiently low temperatures since the coupling constants decays logarithmically, $\alpha_\Lambda\propto1/\ln\left(\frac{\Lambda_0}{T}\right)$. Hence in this case (II), the critical index of Eq. (\ref{PowerLaw}) is $\phi=1$, and we find that, in addition to the exponent, there is nontrivial logarithmic scaling. In Fig.\ref{BcT} we illustrate the asymptotic (\ref{II}) by dashed yellow line originating from $B=T=0$.
The solid blue line originating from the same point represents exact solution of Eq.(\ref{criticalTemp}).

%%%%%%%%%%%%%%%%
%%%%%%%%%%%%%%%%%%%%%%%%%%%%

Finally we consider the BEC case (III), $p<p_c$. In this case only the $\omega_{\bm k}^-$ dispersion branch is critical, $\omega_{\bm k}^-\approx \frac{k^2}{2\Delta_0}$, where $\Delta_0=B_0$ is the gap at $B=0$. The other two branches are gapped. Calculation of $\Sigma_T$ gives $\Sigma_T=\alpha_\Lambda\frac{\zeta(3/2)}{\pi\sqrt{2\pi}}\sqrt{\Delta_0}T^{3/2}$, where $\zeta$ is Riemann's $\zeta$-function. Hence, using Eq. (\ref{criticalTemp}) we find 
\begin{align}
\label{III}
{\text (III):}\hspace{0.2cm} \frac{\delta B_c}{\Delta_0}&=\alpha_\Lambda\frac{\zeta(3/2)}{(2\pi)^{\frac{3}{2}}}\left(\frac{T}{\Delta_0}\right)^{3/2}&&{\text at}\hspace{0.2cm} \delta B_c\ll \Delta_0.
\end{align}
As expected the critical index in Eq. (\ref{PowerLaw}a) is $\phi=3/2$. To understand the region of validity of Eq.(\ref{III}) we compare with TlCuCl$_3$ data \cite{TanakaBc, Oosawa, Shindo}. The value of the gap at $T=p=B=0$ is $\Delta_0=m_{\Lambda,\pm}=0.64$meV \cite{Ruegg2008}. The BEC critical field for $T=p=0$ is $B_0=4.73$ T \cite{Yamada}. Hence, we obtain the $g$-factor, which is defined as $B\to g\mu_B B$, $g=2.35$ \cite{comment}.  In Fig. \ref{BcT} the dashed yellow line originating from  $B_0=4.73$T shows $B_{BEC}$ versus $T$ at $p=0$ calculated with Eq. (\ref{III}). The value of the coupling constant in this equation is obtained from Eq.'s (\ref{coupling}) and (\ref{params}), $\alpha_\Lambda/(8\pi)=\alpha_{\Delta_0}/(8\pi)=0.169$. Experimental data \cite{TanakaBc, Oosawa, Shindo} are shown by circles. We see that Eq. (\ref{III}) is valid only at $T\leq1$K.

There are two physical effects accounted in (\ref{criticalTemp}), but neglect in (\ref{III}). These are (i) the influence of the non-critical (gapped) modes $\omega_{\bm k}^+,\omega_{\bm k}^0$; (ii) the logarithmic running of $\alpha_\Lambda$. To illustrate the importance of non-critical modes, the dashed maroon line originating from $4.73$ T in Fig. \ref{BcT} shows solution solution of Eq. (\ref{criticalTemp}) with account of all three modes, but with fixed coupling constant $\alpha_{\Delta_0}/(8\pi)=0.169$. Finally, the solid blue line originating from $4.73$ T shows solution of (\ref{criticalTemp}) with account of both (i) and (ii). Agreement with experiment is remarkable. We stress that there is no fitting in the theoretical curve. The set of parameters (\ref{params}) was determined in Ref. \cite{Asymptotic} from data unrelated to magnetic field. To be consistent with this set when generating the solid blue and dashed maroon curves in Fig. \ref{BcT} we use the same line broadening as in \cite{Asymptotic}, $\omega_{\bm k}^\sigma\to\sqrt{{\bm k}^2+m_{\Lambda,\sigma}^2+\Gamma_T^2}+\sigma B$, $\Gamma_T=\xi T$, $\xi=0.15$.

Regimes (I) and (II) have never been considered before.
On the other hand, the BEC regime (III) has been considered in a number of publications
using the Hartree-Fock-Popov approximation for hard core bosons, from which simple $T^{3/2}$ dependence is predicted.
Our conclusion is that such an approximation is only valid at vanishingly small temperatures
and the region of validity shrinks to zero upon approaching the critical pressure QCP.
This is illustrated in Fig. \ref{BcT} by lines originating from  points $B_0=4.73$T and
$B_0=2.36$T at $T=0$.
Our exact theoretical solutions (blue solid lines) differ from the simple $T^{3/2}$ dependence 
(dashed yellow) due to two effects; influence the non-critical excitations, and the running of the 
coupling constant. Both effects are governed by the magnetic quantum critical point $(p_c,0,0)$ and cannot be accounted within a hard core boson model;
whether it be Hartree-Fock-Popov approximation or even an exact solution.
Including these effects, the present analysis resolves the long standing problem of the BEC critical exponent, which has been consistently reported at higher value; $3/2\leq\phi\leq2.3$ \cite{BECreview,Shiramura,Kato,Ishii,TanakaBc, Wessel,NohadaniQMC}.

%%%%%%%%%%%%%%%%
\begin{figure}[t!]
{\includegraphics[width=0.238\textwidth,clip]{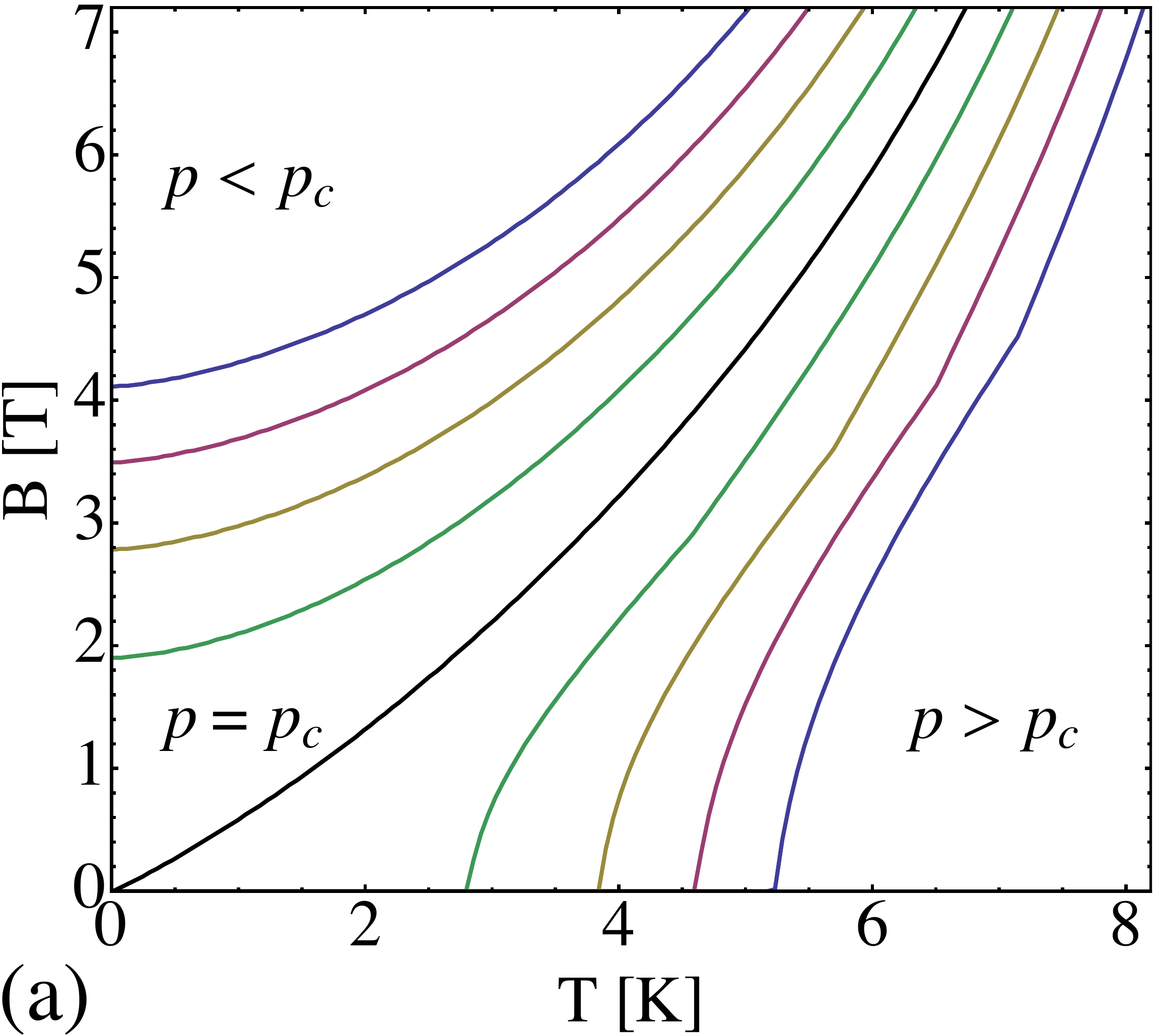}}
{\includegraphics[width=0.238\textwidth,clip]{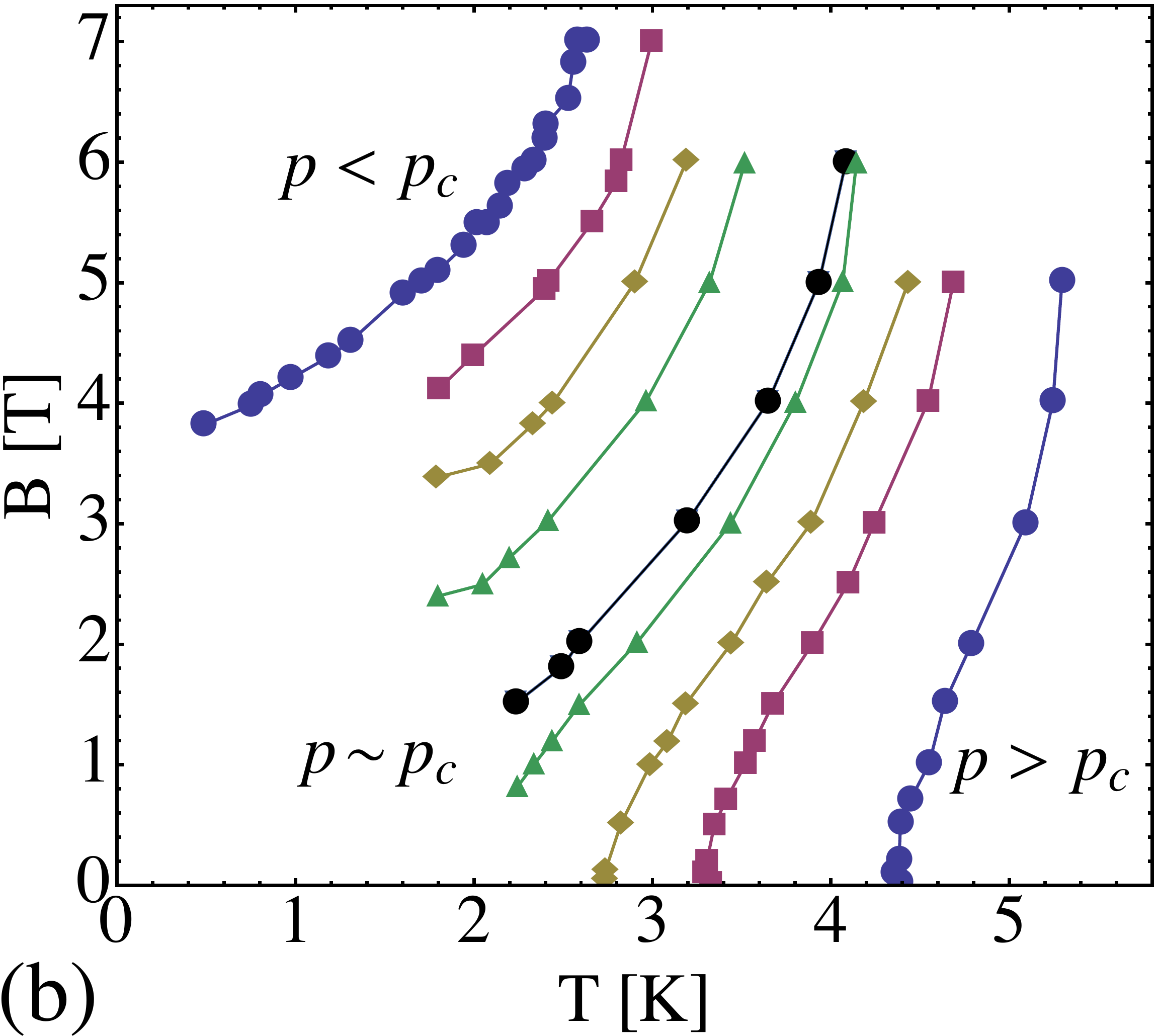}}
\caption{Multiple universalities: Various curves show the critical field $B_c(T)$ at various pressures ranging $p<p_c$, $p=p_c$ to $p>p_c$. (a) Solutions to (\ref{criticalTemp}) with parameters for TlCuCl$_3$. (b) Data for quantum antiferromagnet CsFeCl$_3$ \cite{CsFeCl3}.
 }
\label{CriticalBTexp}
\end{figure}
%%%%%%%%%%%%%%%%%%%%%%%%%%%%

The existence of three critical exponents $\phi=3/2$, $1$ and $1/2$, and even logarithmic 
corrections to these exponents, is a readily testable result and constitutes our most 
important prediction for experiment.
Figure \ref{BcT} provides predictions directly for TlCuCl$_3$. 
In Figure \ref{CriticalBTexp}a  we plot the predicted critical field in TlCuCl$_3$
vs temperature at various pressures.
For comparison in Fig. \ref{CriticalBTexp}b we present  a similar experimental
plot for quantum antiferromagnet CsFeCl$_3$ published very recently \cite{CsFeCl3}. 
Unfortunately we cannot perform exact
quantitative calculations (including all pre-factors) for CsFeCl$_3$. 
Existing data for this compound are not sufficient to perform analysis 
similar to \cite{Asymptotic} for TlCuCl$_3$.  However, the data  \cite{CsFeCl3} supports the 
proposed multiple critical exponent theory.

In summary, employing a quantum field theoretic approach,
our work predicts multiple critical exponents, and their corresponding logarithmic
corrections, on the pressure, magnetic field and temperature - phase
digram for 3D quantum antiferromagnets in vicinity of the
quantum critical point.
For TlCuCl$_3$ we demonstrate remarkable agreement with existing data, and provide quantitative predictions for future experiments.
We also resolve the long standing problem relating to the observed critical exponent
in Bose-Einstein condensation of magnons.

\acknowledgements{
We thank Christian R\"uegg for a very important, stimulating discussion
and Yaroslav Kharkov for drawing our attention to Ref. \cite{CsFeCl3}. 
The work has been supported by the Australian Research Council grant DP160103630.}


\begin{thebibliography}{99}
\bibitem{comment} Due to a small anisotropy, the g-factor slightly depends on the direction of magnetic field with respect to the crystal axes. We use BEC data with magnetic field directed as per experiment.
\end{thebibliography}
\end{document}